\documentclass[conference]{IEEEtran}
\IEEEoverridecommandlockouts
\usepackage{cite}
\usepackage{booktabs}

\usepackage{amsmath,amssymb,amsfonts}
\usepackage{algorithmic}
\usepackage{graphicx}
\usepackage{textcomp}
\usepackage{xcolor}
\def\BibTeX{{\rm B\kern-.05em{\sc i\kern-.025em b}\kern-.08em
    T\kern-.1667em\lower.7ex\hbox{E}\kern-.125emX}}
\begin{document}

\title{A Feature-Driven Framework for Software Fault Prediction\\
}

\author{\IEEEauthorblockN{Ahmad Nauman Ghazi}
\IEEEauthorblockA{\textit{Department of Software Engineering} \\
\textit{Blekinge Institute of Technology}\\
Karlskrona, Sweden \\
nauman.ghazi@bth.se}
\and
\IEEEauthorblockN{Nagajyothi Devarapalli}
\IEEEauthorblockA{\textit{Department of Software Engineering} \\
\textit{Blekinge Institute of Technology}\\
Karlskrona, Sweden \\
nade22@student.bth.se}
\and
\IEEEauthorblockN{Ashir Javeed}
\IEEEauthorblockA{\textit{Department of Computer Science} \\
\textit{Blekinge Institute of Technology}\\
Karlskrona, Sweden \\
ashir.javeed@bth.se}
\and
\IEEEauthorblockN{Sadi Alawadi}
\IEEEauthorblockA{\textit{Department of Computer Science} \\
\textit{Blekinge Institute of Technology}\\
Karlskrona, Sweden \\
sadi.alawadi@bth.se}
\and
\IEEEauthorblockN{Fahed Alkhabbas}
\IEEEauthorblockA{\textit{Dept. of Computer Science and Media Technology} \\
\textit{Malmö University}\\
Malmö, Sweden \\
fahed.alkhabbas@mau.se}
\and
\IEEEauthorblockN{Khalid AlKharabsheh}
\IEEEauthorblockA{\textit{Department of Software Engineering} \\
\textit{Al-Balqa Applied University}\\
Al-Salt, Jordan \\
khalidkh@bau.edu.jo}
}

\maketitle

\begin{abstract}
Software fault prediction (SFP) is a critical task in software engineering, enabling early identification of faults in modules to improve software quality and reduce maintenance costs. This research investigates the combined effects of feature selection and parameter tuning on the performance of machine learning (ML) models for SFP. This study evaluates the interaction between feature selection methods, including correlation-based feature selection (CFS), recursive feature elimination (RFE), mutual information (MI), and L1 regularization, where hyperparameter tuning techniques such as grid search, randomized search, and genetic algorithm (GA) are used for optimization of ML algorithms, including random forest (RF), logistic regression (LR), and support vector machines (SVM) for optimized fault prediction performance. The combined application of CFS and GA yielded the highest accuracy, achieving 88.40\% with RF, representing an improvement of 18\% over baseline models without feature selection or tuning. Feature selection reduced dimensionality and identified critical attributes such as weighted methods per Class (WMC) and coupling between objects (CBO), while iterative parameter tuning optimized model alignment to these feature sets. Notably, the proposed methods demonstrated robustness, with minimal cross-validation variability (±1.0\%), and efficiency, reducing training times in univariate methods such as L1 regularization.
\end{abstract}


\section{Introduction}
One of the biggest obstacles in the software development process is handling the faults. It is important to realize that faults in software development can happen at different levels, from complicated design issues to syntactic errors. Researchers agree that the software testing phase of the software development life cycle (SDLC) is an expensive phase for making corrections~\cite{song2006software}. Early detection of faults in the software development process is essential, particularly during initial stages, such as design and coding, which significantly reduces the time and cost associated with fixing. Throughout the SDLC, software developers ensure trustworthy decision-making by accurately estimating the number of defects. Because of the later identifying and fixing faults following software development activities is costly and time-consuming~\cite{ha2019experimental, selvi2022fault}. The objective of software fault prediction is to detect faulty software modules before the testing phase by using structural characteristics of the software system. This method with three steps, called software fault-prone module classification, is used to predict fault-prone modules for the next release of software. The first step involves collecting historical data or data sets that contain information such as software metrics and fault data of previous releases of similar software projects. The second step is to train the model using the data sets, and the third step is to classify the modules. If the fault is reported during system tests, that module's fault data is marked as 1 (faulty), otherwise 0 (non-faulty).

The idea about the number of fault predictions is attractive because it provides the probability that a certain number of faults will occur in the given software. This type of fault prediction method is useful to narrow down the testing efforts to the software modules having a large number of faults~\cite{rathore2017empirical}. The software metrics, which are also features of the software, reflect the characteristics of software modules. In addition, features or metrics may be more relevant than others, and some may be redundant or irrelevant~\cite{jimoh2018promethee}. The feature selection method improves the quality of the data by reducing the high dimensionality data~\cite{javeed2025predicting}, which consequently improves the predictive performance of ML models. Existing research has shown that irrelevant features, along with redundant features, severely affect the accuracy~\cite{balogun2019performance, javeed2023predicting}.

In parallel, the SFP classifiers are characterized by some configurable parameters, called hyperparameters, that need to be optimized to ensure better performance. The classifiers sometimes do not perform well when default parameters or settings are used for prediction~\cite{khan2020hyper}. The process of tuning these parameters of classifiers or models to find the best hyperparameter value is known as hyperparameter tuning. The classifiers have different parameters that need to be optimized~\cite{tantithamthavorn2018impact}. Hyperparameters significantly impact the performance and efficiency of ML models, and optimal tuning is essential to achieve the best trade-off between bias and variance, as well as maximize predictive accuracy.
Despite these advancements, most studies in software fault prediction tend to focus on either feature selection or hyperparameter tuning as separate optimization tasks. While these techniques have shown promise when used independently, the lack of research on their combined application leaves a critical gap in the pursuit of more efficient and reliable software fault prediction methods. To address this gap, given study proposes an integrated framework that systematically combines hyperparameter tuning and feature selection to optimize ML models for software fault prediction. Furthermore, the key highlights of the given study are as follows:

\begin{itemize}

	\item Integrating feature selection with hyperparameter tuning reduced overfitting and improved test accuracy and cross-validation consistency, especially for RF and SVM.
	
	\item The impact varied by model: RF benefited most from feature interactions, SVM from kernel tuning, and LR showed moderate gains with sparse features.
	
	\item CFS achieved the highest accuracy (88.40\% with RF + GA) by preserving multivariate interactions and reducing redundancy.
	
	\item GA delivered the strongest hyperparameter optimization, particularly when combined with CFS, improving model accuracy and robustness.
	
	\item Randomized search offered a fast, cost-effective alternative, achieving competitive performance with much lower computational expense.
\end{itemize}
 
\section{Related Work}
Researchers have recently expressed a strong interest in using ML approaches to predict software faults. They used ML algorithms to accurately predict software faults. Alsghaier and Akour~\cite{alsghaier2020software} presented a framework based on the genetic algorithm (GA) with support vector machine (SVM) classifier and particle swarm algorithm for software fault prediction. The developed framework provided exceptional results on small-scale and large-scale data sets. AlShaikh and Elmedany~\cite{alshaikh2021estimate} examined the performance of ML algorithms to predict defects in software using the Weka tool. For experimental results, the datasets were used from the PROMISE repository. According to their study, random forest and artificial neural networks have the highest accuracy for predicting defects. 

In another study, Awotunde et al.~\cite{awotunde2022feature} presented a hybrid model using a feature selection-enabled Extreme Gradient Boost (XGB) classifier for software fault prediction. The cleaned NASA MDP datasets were used for the evaluation of the proposed model based on several metrics. Another study by Alkharabsheh et al.~\cite{ALKHARABSHEH2022106736} used 28 ML algorithms, examining their efficiency in defect detection. The experiments were applied using a balanced dataset of 24 Java projects. Well-known performance indicators were computed for ML algorithm comparisons (MCC, ROC, Kappa, precision, and F1-score). The results denoted Cat Boost exceeding the other algorithms, although most of them were efficient in defect detection.
Further, Ali et al.~\cite{ali2023software} studied software fault prediction using ML models. Their study suggested that experimental data provided by academicians promises improved accuracy in software fault prediction. They surveyed a list of notable research publications that have used ML approaches to anticipate software flaws. The most prevalent methods are NB (Naive Bayes), RF (Random Forest), and SVM (Support Vector Machine). Out of these models, they reported that Naive Bayes performs better, followed by RF. Another study~\cite{ali2023analysis} analyzed feature selection methods in software fault prediction using ML models. According to their work, a key factor in the success of software fault prediction is selecting relevant features and reducing data dimensionality. Feature selection methods contribute by filtering out the most critical attributes from a plethora of potential features. These methods have the potential to significantly improve the accuracy and efficiency of fault prediction models.

Nandeesh and Mehta conducted a comparative study on supervised ML models for software fault detection where they analyzed the performance of five ML models, including LR, Gaussian Naive Bayes (GNB), K-Nearest Neighbors (KNN), Quadratic Discriminant Analysis (QDA), and Linear Discriminant Analysis (LDA). The performance of ML models was assessed on evaluation metrics such as precision, recall, F1-score, and accuracy. Results of their study indicate that KNN surpasses the other models~\cite{nandeesh2024comparative}. 

Although prior studies have applied ML models, hybrid techniques, and feature selection methods for software fault prediction with promising results, most works treat feature selection and model parameter optimization independently or only partially. The lack of a unified framework that jointly performs optimal feature selection and hyperparameter optimization across ML models remains a research gap, which motivates the present study.

\section{Methodology}
The research aims to optimize the ML model's performance by selecting the best features (through feature selection) and tuning the model's hyperparameters (via search optimization techniques) to achieve high accuracy, computational efficiency, and effective resource utilization. Figure~\ref{flowchart} presents the workflow of the proposed study for effective software fault prediction. From Figure~\ref{flowchart}, it is evident that the proposed framework consists of seven steps. The first step is based on dataset collection, while the second step presents the techniques and methods for data preprocessing used in the given framework. The third step focuses on data partitioning using a cross-validation (K = 10) scheme for training and testing the ML models. Step four tackles the problem of class imbalance in the dataset using adaptive synthetic sampling (ADASYN). Step five of the given framework deals with feature selection methods for selecting the most significant features from the dataset for classification. Step six is based on the optimization and fine-tuning of the ML models. In step seven, the optimized and fine-tuned ML models are tested on the test dataset. The description all seven steps of proposed framework are given as follows:

\begin{figure*}[!t]
	\centering
	\includegraphics[width=0.80\textwidth]{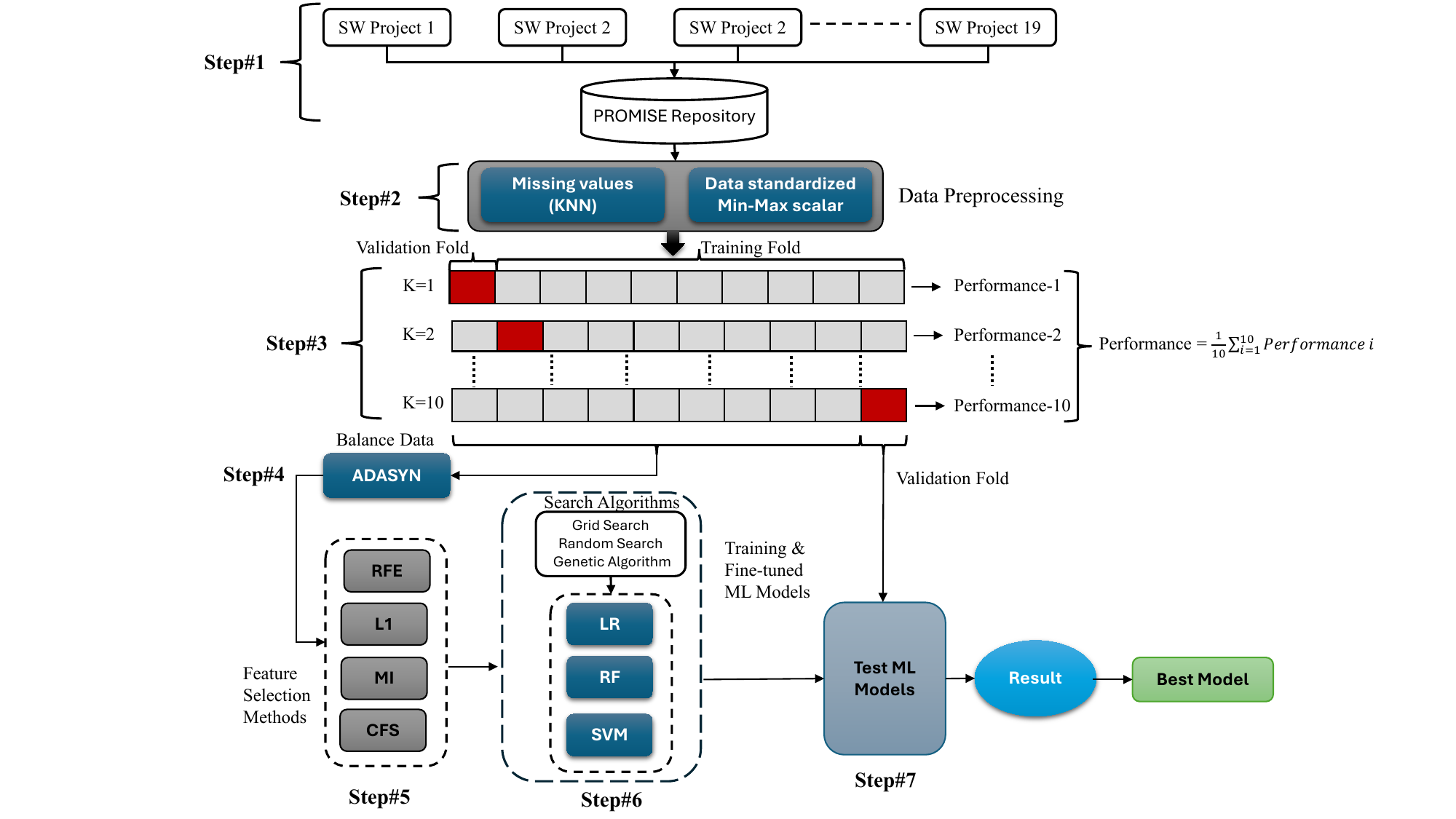}
	\caption{Working of the proposed framework}
	\label{flowchart}
	\end{figure*}

\subsection{Data Cleaning and Pre-processing}
The effectiveness of fault prediction models depends on the quality of the data used~\cite{hall2012state}. To ensure that each record is unique, we identified and removed any duplicates. Duplicate software modules may arise from repeated code or multiple versions of the same project. By ensuring that each module is represented only once, we maintain the integrity and impartiality of the dataset.

Several crucial indicators need to be included, likely due to incomplete data collected during software development.  We used the K-Nearest Neighbors (KNN) method for imputation to estimate and fill in the missing variables.  We standardized the dataset values using the min-max scaling method to ensure optimal performance for ML algorithms. To prevent features with larger scales from dominating the model training process and to facilitate smoother convergence for optimization techniques, this scaling strategy adjusts all features to fall within a consistent range of 0 to 1.

\subsection{Data splitting using K-fold Stratified Cross Validation}
Cross-validation is a technique used to validate a model on different subsets of the data, ensuring that the models generalize well to unseen data and are not overfitting to the training set. Instead of splitting data into just one training and testing set (holdout validation), k-fold stratified cross-validation splits it into k-subsets. The data is split using 10-fold cross-validation, the most commonly used method~\cite{bates2024cross}. The cleaned dataset comprises thousands of valid samples. To rigorously evaluate the performance of the model, a 10-fold stratified cross-validation approach was employed. In this method, the dataset was divided into ten equally sized subgroups (folds) while preserving the class distribution across each fold to address the inherent imbalance in the dataset.
During each iteration of cross-validation, the model was trained on nine folds (90\% of the data) and validated on the remaining fold (10\% of the data). This process was repeated ten times, ensuring that each fold was used as the validation set exactly once. The final performance metrics were averaged across all ten iterations, providing a robust and unbiased estimate of the model's generalization capabilities.
Stratified cross-validation was chosen to ensure that each fold retained the same class distribution as the original dataset. This is particularly important for imbalanced datasets, such as those used in software fault prediction, where faulty modules often represent a minority class. Without stratification, some folds might lack representation of the minority class, leading to misleading evaluation metrics.
\subsection{Feature Selection}
This research focuses on four feature selection methods: Recursive Feature
Elimination (RFE), L1 regularization, Mutual Information (MI), and Correlation-based Feature Selection(CFS). Each technique is applied separately to baseline models, allowing for an evaluation of their performance in feature selection. The goal is to explore the importance of each feature selection method.
\subsubsection{Recursive Feature Elimination (RFE)}
RFE (Recursive Feature Elimination) is an iterative, wrapper-based technique that repeatedly trains a model to assess the importance of various features. In each iteration, it ranks the features by their significance and removes the least important ones. This process continues until a desired subset of features is retained.
RFE evaluates feature importance dynamically by using coefficients in linear models or feature importance scores in tree-based models. In this study, RFE identified and retained critical features that are highly predictive of software faults, including Weighted Methods per Class (WMC), Response for a Class (RFC), Coupling Between Objects (CBO), and Cyclomatic Complexity (CC). In contrast, features like the Number of Children (NOC) and Measure of Aggregation (MOA) were eliminated due to their weaker contributions.
\subsubsection{L1 Regularization}
L1 Regularization imposes a penalty on the absolute value of feature coefficients in regression models, effectively reducing some coefficients to zero. This sparsity mechanism helps select a subset of features while discarding those that are less relevant. In this study, L1 Regularization identified features such as wmc, ce, and Average Cyclomatic Complexity (avg\_cc) as positively weighted contributors. It also included features like CBO and loc, despite their negative relationships with the target variable. Features with coefficients of zero, such as noc, Depth of Inheritance Tree (dit), and Measure of Functional Abstraction (mfa), were excluded from consideration.
\subsubsection{Mutual Information(MI)}
Mutual Information (MI) measures the dependency between features and the target variable without assuming linear relationships. This makes it particularly useful for capturing non-linear dependencies. By quantifying the amount of information a feature provides about the target variable, MI identified features such as Weight Method Classes (WMC), Response For Class (RFC), Lines of Code (LOC), and Coupling Between Objects (CBO) as the most informative. In contrast, features like Number of Children (NOC), Afferent Coupling (Ca), and Inheritance Coupling (IC) were excluded due to their weak associations with the target variable.
\subsubsection{Correlation-Based Feature Selection (CFS)}
CFS (Correlation-based Feature Selection) evaluates subsets of features based on their relevance to the target variable and their redundancy with one another. Its goal is to maximize the correlation between selected features and the target variable while minimizing the correlations among the features themselves. In this study, CFS identified a concise yet highly relevant set of features, retaining weight modification complexity (wmc), coupling between objects (cbo), lines of code (loc), response for class (rfc), and cyclomatic complexity (ce), while excluding redundant features such as number of children (noc) and class attributes (Ca). CFS achieved an optimal balance between dimensionality reduction and predictive accuracy by focusing on predictors driven by interactions.

\subsection{Machine learning model}
\subsubsection{Random Forest} 
An effective method for determining whether or not a particular software module is defective is Random Forest (RF). For tasks involving binary classification, it is especially helpful~\cite{agrawallaensemble}. Because it uses an ensemble of several decision trees, it can handle high-dimensional data and resists overfitting, which is one of its main advantages.
The Random Forest algorithm builds multiple decision trees using the following steps~\cite{javeed2024unveiling}:
\begin{itemize}
	\item Bootstrap sampling: Bootstrap sampling involves creating random subsets of the data by sampling with replacement. Each of these subsets is used to train an individual decision tree.
	\item Random Feature Selection: At each split in a tree, a random subset of features is considered, which adds diversity among the trees.
	\item Model Aggregation: After all the trees are trained, predictions are aggregated using majority voting for classification or averaging for regression.
\end{itemize}

These mechanisms help reduce overfitting and enhance the model's ability to generalize to unseen data, making it highly effective for predictive tasks like software fault prediction. By averaging predictions across multiple trees, Random Forest decreases the risk of overfitting, especially in noisy software datasets. Additionally, Random Forest includes a built-in feature to estimate the importance of various metrics, which assists in identifying the most critical indicators for predicting software faults~\cite{karim2017software}.
\subsubsection{Logistic Regression}
Logistic regression is a statistical method used to predict binary outcomes based on a set of independent variables. In the realm of software fault prediction, logistic regression estimates the likelihood that a software module contains a fault, using various software metrics~\cite{rathore2021software, salem2004prediction}. \\\\
The logistic regression model takes a collection of predictor variables $X = (X_1, X_2, \ldots, X_p)$ and predicts the probability $P(Y=1|X)$ that a software module $Y$ is faulty.  The model is defined as below~\cite{salem2004prediction}:

\begin{equation} 
P(Y=1|X) = \frac{1}{1 + e^{-(\beta_0 + \beta_1 X_1 + \beta_2 X_2 + \ldots + \beta_p X_p)}}
\end{equation} 
Where:
\begin{itemize}
	\item $Y$ is the dependent binary variable (1 if the module is faulty, 0 otherwise).
	\item $X_1, X_2, \ldots, X_p$ are the independent variables (software metrics).
	\item $\beta_0, \beta_1, \beta_2, \ldots, \beta_p$ are the coefficients of the model.
\end{itemize}
The logistic function takes a linear combination of input features and transforms it into a value between 0 and 1. This value can be interpreted as the probability that a component is faulty. In logistic regression, the coefficients indicate both the direction and strength of the relationship between each feature and the likelihood of a fault occurring. For example, a positive coefficient for a metric like "lines of code" suggests that higher values of this metric are associated with a greater likelihood of faults.
\subsubsection{Support Vector Machine} 
\noindent The Support Vector Machine (SVM) is a powerful ML algorithm utilized for classification and regression tasks, including the prediction of software faults. Developed by Vladimir Vapnik and Alexey Cortes in 1995, SVM seeks to identify an optimal hyperplane that divides the data into distinct classes while maximizing the margin between them. The working mechanism of SVM is given as follows: 

Hyperplane and Margin: The Support Vector Machine (SVM) algorithm aims to find a hyperplane that maximizes the margin between two classes. In a binary classification problem, the hyperplane can be defined as \cite{rath2022comparative}:
\[
\mathbf{w} \cdot \mathbf{x} + b = 0
\]
where \(\mathbf{w}\) is the weight vector, \(\mathbf{x}\) is the input vector, and \(b\) is the bias term.

Maximizing the Margin: The margin is the distance between the hyperplane and the nearest data points from both classes, known as support vectors. The objective of SVM is to maximize this margin, which can be formulated as a convex optimization problem~\cite{rath2022comparative}: 
\[
y_i (\mathbf{w} \cdot \mathbf{x}_i + b) \geq 1 \quad \forall i
\]
where \(y_i\) is the class label of the \(i\)-th data point.

Kernel Trick: For non-linearly separable data, SVM can transform the input space into a higher-dimensional space using kernel functions. This allows the algorithm to find a linear hyperplane in this new space. Common kernel functions include~\cite{rath2022comparative}:
\begin{itemize}
	\item  Linear Kernel: For linearly separable data.
	\item Polynomial Kernel: For polynomial decision boundaries.
	\item Radial Basis Function (RBF) Kernel: For non-linear data.
\end{itemize}

Random Forest, Logistic Regression, and Support Vector Machine are the three baseline models used in this research to predict software faults. As illustrated in Fig.~\ref{flowchart}, feature selection is carried out using Recursive Feature Elimination (RFE), L1 Regularization, Mutual Information (MI), and Correlation-based Feature Selection (CFS), all of which are combined with K-Fold Cross-Validation (K=10). Model training is then conducted using the selected features, without any hyperparameter tuning at this stage. Each model is evaluated both with and without the feature selection process. Performance metrics such as accuracy, precision, recall, F1 score, and computational efficiency are computed. This sets a baseline for comparison when more advanced techniques, like hyperparameter optimization, are applied later.

\subsection{Hyperparameter Optimization (Search Optimization Techniques)}
The hyperparameters of each model are adjusted using optimization techniques once the baseline models have been evaluated. The methods employed include grid search, randomized search, and genetic algorithms. These techniques systematically explore the parameter space to identify the optimal set of hyperparameters. Hyperparameters are settings defined before training that significantly influence the learning process. Examples include the learning rate, the number of epochs, and algorithm-specific parameters such as the kernel type in Support Vector Machines (SVM) or the tree depth in Random Forest. Adjusting these hyperparameters can have a substantial impact on model performance. This significance is underscored by the work of Tantithamthavorn et al.~\cite{tantithamthavorn2016automated} and Fu et al.~\cite{fu2016tuning}, who highlighted the importance of systematic tuning for achieving optimal results.

Grid Search:
Grid search systematically explores a predefined subset of the hyperparameter space to identify the optimal settings for a learning algorithm~\cite{khan2024machine}. It evaluates each combination using cross-validation, ensuring reliable performance assessment. The combination yielding the best performance metric (e.g., accuracy, precision, recall, F1 score) is selected. Key components include:
\begin{itemize}
	\item Cross-validation: Commonly uses k-fold cross-validation to train and validate the model across different data splits.
	\item Performance Metrics: Metrics such as accuracy, precision, recall, and F1 score guide the selection of optimal hyperparameters.
\end{itemize}

Randomized Search:
Randomized search improves efficiency by sampling from the hyperparameter space rather than exhaustively searching every combination. This approach is advantageous for models with large parameter spaces, offering a balance between computational cost and performance. Random sampling from distributions like uniform or log-uniform is combined with cross-validation to assess performance. The method is especially useful when computational resources are limited.

Genetic Algorithm:
Genetic Algorithm (GA) optimization technique is inspired by the process of natural selection, making them particularly effective for fine-tuning hyperparameters. These algorithms iteratively evolve a population of candidate solutions through the processes of selection, crossover, and mutation to identify optimal parameter settings~\cite{tantithamthavorn2016automated, javed2023localization}. The key steps include:
\begin{itemize}
	\item Selection: This step involves choosing the best-performing individuals to carry on to the next generation.
	\item Crossover: In this stage, selected candidates are combined to produce offspring, which fosters genetic diversity.
	\item Mutation: Random variations are introduced to the solutions to prevent premature convergence and maintain diversity.
\end{itemize}
The fitness function assesses model performance using metrics like accuracy and F1 score across cross-validation folds, helping guide the evolution toward better parameter settings. This method is especially effective in exploring complex, high-dimensional search spaces. Tuning Parameters for Baseline Models is given as:

Random Forest: Key parameters include the number of trees (\textit{n\_estimators}), tree depth (\textit{max\_depth}), minimum samples for splits (\textit{min\_samples\_split}), and leaf nodes (\textit{min\_samples\_leaf}).

Logistic Regression: Parameters include the regularization strength (\textit{C}), penalty type (\textit{l1}, \textit{l2}), and solver (\textit{liblinear}).

SVM: Important hyperparameters include the regularization parameter (\textit{C}), kernel type (e.g., \textit{linear}, \textit{rbf}), and \textit{gamma} for RBF kernels. 
Upon feature selection and hyperparameter tuning, K-fold stratified cross-validation is used to identify the best features and hyperparameters for training the final models.
Every cross-validation iteration after the models are fully trained assesses them on the validation fold.

\section{Dataset Description}
The fault prediction dataset is obtained from the PROMISE Repository, which is a publicly available repository specializing in software engineering research datasets. This dataset is widely used in software fault prediction due to its comprehensive inclusion of software metrics and defect labels, making it reflective of real-world software problems. The study investigates seven open-source software projects, each containing multiple versions, totaling 19 open-source software projects. Detailed information such as version numbers, file class names, and defect labels for every source file are included for each project. Table~\ref{Dataset and Defective Instances} provides a summary of the collected data consisting of the selected software project (Dataset), the total number of instances in the dataset, defective instances in the given dataset, and the rate of defective instances. Each of these open-source projects represented a different software module (or class) with 21 unique features(metrics). 
\begin{table}[h]
	\centering
	\caption{Dataset and Defective Instances}
	\label{Dataset and Defective Instances}
	\begin{tabular}{lccc}
		\toprule
		\textbf{Dataset} & \textbf{Inst.} & \textbf{Defective Inst.} & \textbf{Rate of Defective Inst.} \\
		\midrule
		ant-1.7    & 746 & 166 & 0.223 \\
		camel-1.0  & 340  & 13  & 0.038 \\
		camel-1.2  & 608  & 216 & 0.355 \\
		camel-1.4  & 872  & 145 & 0.166 \\
		camel-1.6  & 966  & 188 & 0.195 \\
		jedit-3.2  & 276  & 90  & 0.331 \\
		jedit-4.0  & 306  & 75  & 0.245 \\
		jedit-4.2  & 368  & 48  & 0.131 \\
		jedit-4.3  & 492  & 11  & 0.022 \\
		log4j-1.0  & 135  & 34  & 0.252 \\
		log4j-1.1  & 110  & 37  & 0.339 \\
		log4j-1.2  & 205  & 189 & 0.922 \\
		lucene-2.0 & 196  & 91  & 0.467 \\
		lucene-2.2 & 247  & 144 & 0.583 \\
		synapse-1.0 & 158 & 21    & 0.132 \\
		synapse-1.2 & 257 & 145   & 0.564 \\
		xalan-2.0   & 724 & 156    & 0.215 \\
		xalan-2.4   &723  & 110 & 0.152\\
		xalan-2.6   & 885 &411	& 0.464\\
		\bottomrule
	\end{tabular}
	
\end{table}

\section{Evaluation metrics}
For every model, metrics for performance are collected, including Accuracy, F1 score, Precision, Recall, and computational efficiency. To evaluate the overall performance of the model, these metrics are averaged over all the folds. Based on these metrics, the final comparison of the models (Random Forest, Logistic Regression, SVM) is performed with and without feature selection, using optimal hyperparameters. It is not enough to rely solely on accuracy (eq.~\ref{accuracy formula})~\cite{medeshetty2025requirements} when evaluating a software fault prediction model, particularly when dealing with imbalanced datasets where faults are a minority~\cite{rathore2022generative, javed2024optimizing}.

\begin{equation} \text{Accuracy} = \frac{TP + TN}{TP + TN + FP + FN}  \label{accuracy formula} \end{equation}

\noindent For example, a model may achieve high accuracy by simply predicting "no-fault" for all modules. To gain a more nuanced understanding of the trade-off between different aspects, it is essential to consider other metrics such as precision (eq.~\ref{precision formula}), recall (eq.~\ref{recall formula}), F1 score (eq.~\ref{f1 score formula}), support, and computational efficiency. These metrics encompass correctly and incorrectly classified instances while also accounting for the time and resource consumption of the model. By doing so, the model can be optimized according to the specific needs of software fault prediction~\cite{kaliraj2024software, sohn2021leveraging}.

Precision indicates the model's ability to avoid false positives, meaning how many predicted faults were actually faults.

\begin{equation} \text{Precision} = \frac{TP}{TP + FP} \label{precision formula} \end{equation}

Recall helps determine how good the model is at identifying actual faults in the software. It measures the proportion of actual faults correctly predicted by the model. 

\begin{equation} \text{Recall} = \frac{TP}{TP + FN} \label{recall formula} \end{equation}

F1 Score balances precision and recall, providing a more comprehensive picture of the model's performance. 

\begin{equation} F1 \text{ score} = 2 \cdot \frac{\text{Precision} \cdot \text{Recall}}{\text{Precision} + \text{Recall}} \label{f1 score formula} \end{equation}

Alongside performance metrics, computational efficiency is crucial for evaluating models, particularly in resource-constrained environments. It involves measuring the time taken for training and testing, as well as the memory and computational resources required. Models with shorter training and testing times are particularly beneficial for real-time or large-scale software fault prediction tasks. To ensure reliable and generalize results, the models were evaluated using 10-fold Stratified Cross-Validation\cite{javeed2025improving}.

\section{Experimental Results}
This section presents a comprehensive evaluation of the selected ML models (LR, RF, SVM) using 10-fold stratified cross-validation, ensuring that class proportions are preserved across all folds. The experimental analysis considers multiple feature selection strategies such as Recursive Feature Elimination (RFE), L1 regularization, Mutual Information (MI), and Correlation-based Feature Selection (CFS) combined with three hyperparameter optimization techniques: Grid Search (GS), Random Search (RS), and Genetic Algorithm (GA). Model performance is assessed using evaluation metrics such as F1-score, precision, recall, and accuracy, as well as training and testing time, providing a balanced view of predictive effectiveness and computational efficiency.

Table~\ref{tab:rfe_optimization} summarizes the results obtained using RFE on balanced data. Among the evaluated models, RF consistently outperforms LR and SVM across all optimization strategies. The highest performance is achieved by RF with GA optimization, attaining an F1-score of 0.87, precision of 0.88, recall of 0.86, and accuracy of 87.41\%, indicating strong classification capability and balanced error handling.

For LR and SVM, GA optimization also yields superior results compared to GS and RS, demonstrating improved convergence toward optimal hyperparameters. Although LR shows lower computational cost, its predictive performance remains slightly inferior to RF, while SVM provides competitive results at the expense of higher testing time.
\begin{table}[t]
	\caption{Performance of ML models using RFE with hyperparameter optimization on balanced data}
	\label{tab:rfe_optimization}
	\centering
	\renewcommand{\arraystretch}{1.15}
	\begin{tabular}{llcccccc}
		\toprule
		\textbf{M} & \textbf{Opt.} & \textbf{F1} & \textbf{Pr.} & \textbf{Re.} &
		\textbf{Acc.(\%)} & \textbf{Train(s)} & \textbf{Test(s)} \\
		\midrule
		RF  & GS & 0.85 & 0.86 & 0.84 & 84.12 & 10.3 & 15.8 \\
		RF  & RS & 0.86 & 0.87 & 0.85 & 85.25 & 8.7  & 12.4 \\
		RF  & GA & 0.87 & 0.88 & 0.86 & 87.41 & 9.4  & 14.7 \\

		LR  & GS & 0.82 & 0.84 & 0.81 & 81.74 & 4.2 & 6.8 \\
		LR  & RS & 0.83 & 0.85 & 0.82 & 83.21 & 3.9 & 6.1 \\
		LR  & GA & 0.84 & 0.85 & 0.83 & 83.92 & 5.0 & 7.4 \\

		SVM & GS & 0.82 & 0.84 & 0.81 & 82.47 & 6.2 & 9.1 \\
		SVM & RS & 0.84 & 0.85 & 0.83 & 83.85 & 5.8 & 8.3 \\
		SVM & GA & 0.85 & 0.86 & 0.85 & 84.62 & 6.7 & 9.9 \\
		\bottomrule
	\end{tabular}
\end{table}

The performance outcomes using L1 regularization are presented in Table~\ref{tab:l1_optimization}. Overall, models trained with L1-based feature selection exhibit lower F1-scores and precision values compared to RFE and other methods. This behavior suggests that L1 regularization may excessively penalize informative features in this dataset.

Despite this limitation, RF with RS optimization achieves the highest accuracy of 86.32\%, although its F1-score remains moderate at 0.75, reflecting a trade-off between recall and precision. LR and SVM models show relatively stable recall values (around 0.82–0.83), but reduced precision, indicating a higher false-positive rate.

\begin{table}[t]
	\caption{Performance of ML models using L1 regularization with hyperparameter optimization}
	\label{tab:l1_optimization}
	\centering
	\scriptsize
	\renewcommand{\arraystretch}{1.05}
	\begin{tabular}{llcccccc}
		\toprule
		\textbf{M} & \textbf{Opt.} & \textbf{F1} & \textbf{Pr.} & \textbf{Re.} &
		\textbf{Acc.} & \textbf{Train(s)} & \textbf{Test(s)} \\
		\midrule
		RF  & GS & 0.75 & 0.71 & 0.82 & 80.16 & 9.4 & 14.7 \\
		RF  & RS & 0.75 & 0.69 & 0.82 & 86.32 & 8.6 & 12.2 \\
		RF  & GA & 0.74 & 0.68 & 0.82 & 82.06 & 12.0 & 18.0 \\

		LR  & GS & 0.75 & 0.84 & 0.70 & 81.26 & 3.8 & 6.4 \\
		LR  & RS & 0.73 & 0.71 & 0.82 & 80.71 & 3.6 & 5.9 \\
		LR  & GA & 0.75 & 0.71 & 0.83 & 80.73 & 4.5 & 7.0 \\

		SVM & GS & 0.74 & 0.68 & 0.82 & 80.20 & 5.9 & 8.5 \\
		SVM & RS & 0.74 & 0.68 & 0.83 & 80.86 & 5.6 & 7.9 \\
		SVM & GA & 0.75 & 0.68 & 0.82 & 83.92 & 6.5 & 9.4 \\
		\bottomrule
	\end{tabular}
\end{table}
Table~\ref{tab:mi_optimization} reports the results obtained using Mutual Information for feature selection. Compared to L1 regularization, MI provides a noticeable improvement in all metrics. RF with GA optimization again delivers the best performance, achieving an F1-score of 0.87, recall of 0.88, and accuracy of 87.00\%.

Both LR and SVM benefit from GA-based optimization, showing consistent improvements in F1-score and accuracy. These results indicate that MI is effective in identifying informative features while maintaining model generalization across cross-validation folds.
\begin{table}[t]
	\caption{Performance of ML models using mutual information with hyperparameter optimization}
	\label{tab:mi_optimization}
	\centering
	\scriptsize
	\renewcommand{\arraystretch}{1.05}
	\begin{tabular}{llcccccc}
		\toprule
		\textbf{M} & \textbf{Opt.} & \textbf{F1} & \textbf{Pr.} & \textbf{Re.} &
		\textbf{Acc.} & \textbf{Train(s)} & \textbf{Test(s)} \\
		\midrule
		RF  & GS & 0.84 & 0.85 & 0.83 & 84.30 & 10.1 & 14.9 \\
		RF  & RS & 0.85 & 0.86 & 0.84 & 83.54 & 8.5  & 12.7 \\
		RF  & GA & 0.87 & 0.86 & 0.88 & 87.00 & 11.7 & 17.8 \\

		LR  & GS & 0.81 & 0.82 & 0.80 & 80.98 & 4.3 & 6.9 \\
		LR  & RS & 0.82 & 0.83 & 0.81 & 81.67 & 4.0 & 6.2 \\
		LR  & GA & 0.83 & 0.84 & 0.82 & 83.04 & 4.9 & 7.3 \\

		SVM & GS & 0.81 & 0.82 & 0.80 & 81.56 & 6.3 & 9.3 \\
		SVM & RS & 0.82 & 0.83 & 0.81 & 82.34 & 5.9 & 8.7 \\
		SVM & GA & 0.83 & 0.84 & 0.82 & 83.21 & 6.8 & 9.6 \\
		\bottomrule
	\end{tabular}
\end{table}

The best overall performance is observed with Correlation-based Feature Selection (CFS), as shown in Table~\ref{tab:cfs_optimization}. In particular, RF combined with GA optimization achieves the highest scores across all evaluated metrics, with an F1-score of 0.88, precision of 0.89, recall of 0.87, and accuracy of 88.40\%. These results highlight the effectiveness of CFS in reducing feature redundancy while preserving discriminative power.

LR and SVM also demonstrate consistent improvements under CFS, with GA optimization outperforming GS and RS. Although GA incurs higher training and testing time, the performance gains justify its computational overhead in scenarios where accuracy is critical.

\begin{table}[t]
	\caption{Performance of ML models using correlation-based feature selection with hyperparameter optimization}
	\label{tab:cfs_optimization}
	\centering
	\scriptsize
	\renewcommand{\arraystretch}{1.05}
	\begin{tabular}{llcccccc}
		\toprule
		\textbf{M} & \textbf{Opt.} & \textbf{F1} & \textbf{Pr.} & \textbf{Re.} &
		\textbf{Acc.} & \textbf{Train(s)} & \textbf{Test(s)} \\
		\midrule
		RF  & GS & 0.85 & 0.86 & 0.85 & 85.76 & 9.5 & 13.8 \\
		RF  & RS & 0.87 & 0.88 & 0.86 & 86.90 & 8.4 & 11.9 \\
		RF  & GA & \textbf{0.88} & \textbf{0.89} & \textbf{0.87} & \textbf{88.40} & 12.5 & 17.6 \\

		LR  & GS & 0.82 & 0.83 & 0.82 & 82.12 & 3.9 & 6.3 \\
		LR  & RS & 0.83 & 0.84 & 0.83 & 83.04 & 3.7 & 5.8 \\
		LR  & GA & 0.84 & 0.85 & 0.84 & 84.15 & 4.7 & 6.9 \\

		SVM & GS & 0.83 & 0.84 & 0.82 & 82.88 & 6.0 & 8.8 \\
		SVM & RS & 0.84 & 0.85 & 0.83 & 83.76 & 5.7 & 8.2 \\
		SVM & GA & 0.85 & 0.86 & 0.85 & 85.12 & 6.9 & 9.8 \\
		\bottomrule
	\end{tabular}
\end{table}

Table~\ref{tab:tuning_comparison} compares model accuracies obtained with and without hyperparameter tuning. Across all models, hyperparameter optimization significantly improves accuracy, with GA and RS generally outperforming GS. For example, RF accuracy increases from 70.21\% without tuning to 86.32\% with RS and 82.06\% with GA, underscoring the importance of systematic tuning in improving generalization performance.

\begin{table}[t]
	\caption{Comparison of model accuracies with and without hyperparameter tuning}
	\label{tab:tuning_comparison}
	\centering
	\scriptsize
	\renewcommand{\arraystretch}{1.05}
	\begin{tabular}{lcccc}
		\toprule
		\textbf{Model} & \textbf{GS} & \textbf{RS} & \textbf{GA} & \textbf{No-Tune} \\
		\midrule
		RF  & 80.16 & 82.06 & 86.32 & 70.21 \\
		LR  & 81.26 & 81.27 & 83.26 & 60.47 \\
		SVM & 81.53 & 81.53 & 82.32 & 63.05 \\
		\bottomrule
	\end{tabular}
\end{table}

Finally, Table~\ref{tab:overfitting} presents baseline results without feature selection or hyperparameter tuning. A clear discrepancy between training and testing accuracy is observed for all models, particularly for RF, indicating overfitting. These findings confirm that the combined use of feature selection, hyperparameter optimization, and stratified cross-validation is essential for achieving robust and generalizable models.

\begin{table}[t]
	\caption{Model performance without feature selection or hyperparameter tuning}
	\label{tab:overfitting}
	\centering
	\scriptsize
	\renewcommand{\arraystretch}{1.05}
	\begin{tabular}{lcccc}
		\toprule
		\textbf{Model} & \textbf{Train (s)} & \textbf{Pred. (s)} & \textbf{Train Acc.} & \textbf{Test Acc.} \\
		\midrule
		RF  & 7.0 & 0.18 & 81.52 & 70.21 \\
		LR  & 6.0 & 0.15 & 73.67 & 60.47 \\
		SVM & 2.2 & 0.49 & 75.24 & 63.05 \\
		\bottomrule
	\end{tabular}
	
	\vspace{1mm}
	\footnotesize Train: Training time (s), Pred.: Prediction time (s), Acc.: Accuracy (\%).
\end{table}

Overall, the results show that Correlation-based Feature Selection combined with Genetic Algorithm optimization delivers the highest and most consistent performance across all evaluation metrics. Random Forest outperforms Logistic Regression and SVM under all feature selection strategies, while 10-fold stratified cross-validation and hyperparameter tuning play a crucial role in ensuring robust generalization and improved predictive accuracy.

\begin{figure}
	\centering
	\includegraphics[width=0.40\textwidth]{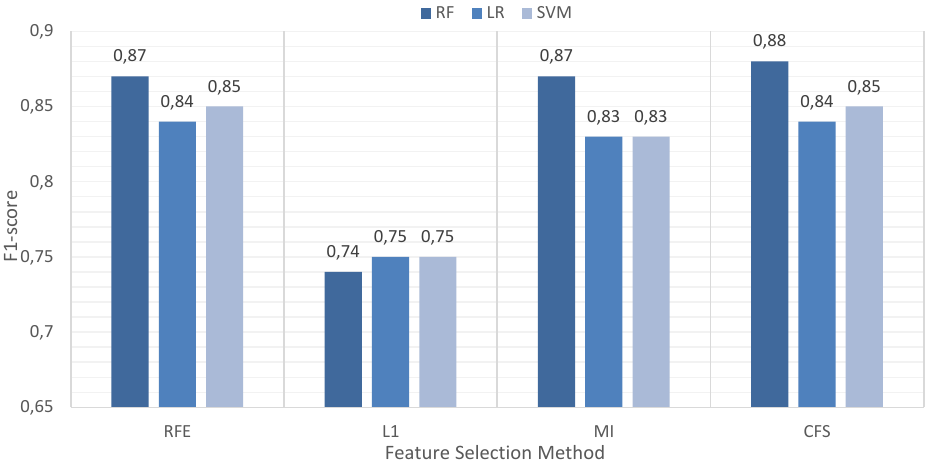}
	\caption{ML models performance comparison based on F1 score}
	\label{F1_ML_Results_Comparison}
\end{figure}

Furthermore, we analyzed the selected ML model performance based on feature selection methods. Figure~\ref{F1_ML_Results_Comparison} presents the results of ML models based on F1-score. It can be observed from Figure~\ref{F1_ML_Results_Comparison} that the RF model achieved the highest F1-score in comparison to the rest of the ML models based on feature selection methods.

\begin{figure}
	\centering
	\includegraphics[width=0.40\textwidth]{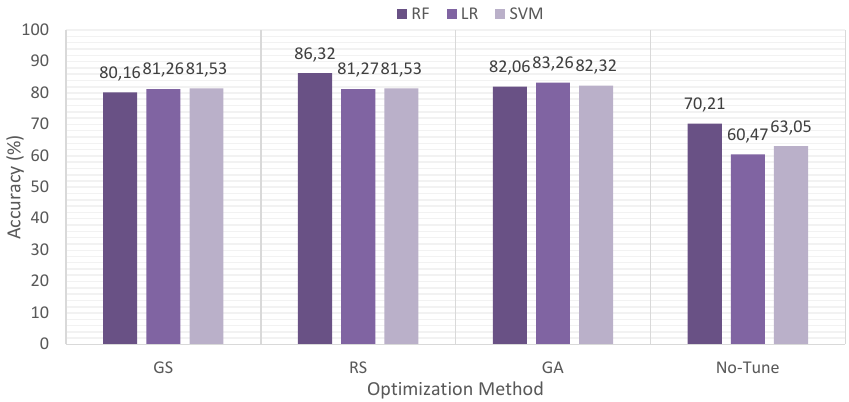}
	\caption{Optimized ML models performance comparison based on accuracy}
	\label{OP_ML_Results_Comparison}
\end{figure}

Figure~\ref{OP_ML_Results_Comparison} presents the impact of the optimization method on the performance of ML models in terms of accuracy. It can be observed from Figure~\ref{OP_ML_Results_Comparison} that ML models perform significantly lower when hyperparameters are not tuned and the ML model is not optimized.

\begin{figure}
	\centering
	\includegraphics[width=0.40\textwidth]{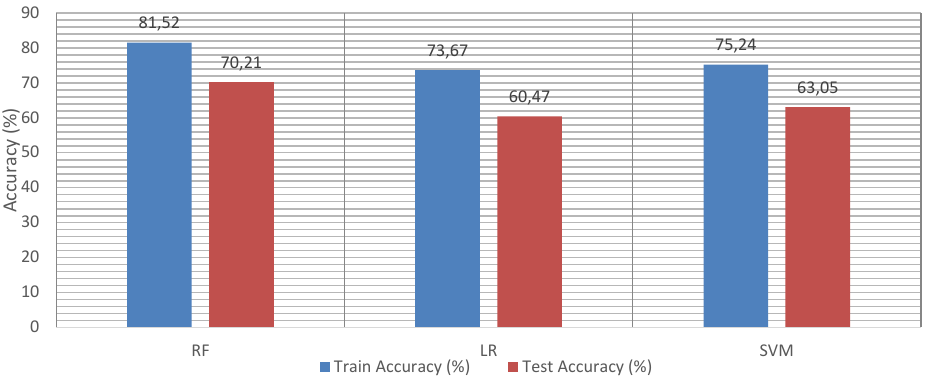}
	\caption{Over-fitting analysis of ML models}
	\label{Overfitting_ML_Analysis}
\end{figure}

Moreover, Figure~\ref{Overfitting_ML_Analysis} shows the training and testing accuracies of three machine learning models, namely RF, LR, and SVM. For all models, the training accuracy is higher than the corresponding testing accuracy. The highest training accuracy (81.52\%) and testing accuracy (70.21\%) are achieved by RF, followed by SVM (75.24\% training, 63.05\% testing) and LR (73.67\% training, 60.47\% testing). This comparison highlights the difference in model performance between seen (training) and unseen (testing) data, indicating the presence of over-fitting.

\section{Conclusion}
The given work examines the combined effects of feature selection methods along with hyperparameter optimization on ML models for software fault prediction. The study results illustrated that feature selection and tuning strategies significantly improve predictive accuracy and computational cost of ML models. Multivariate feature selection, particularly correlation-based feature selection (CFS), effectively preserved feature interactions, enabling optimized RF through GA, which achieved the highest accuracy of 88.40\% with low cross-validation variance.

In contrast, univariate methods such as mutual information and L1 regularization emphasized efficiency and, when combined with randomized search, achieved competitive accuracy with substantially reduced training time, making them suitable for time-sensitive applications. The findings confirm that optimal configurations depend on model characteristics, with ensemble and kernel-based models benefiting from multivariate selection and iterative tuning, while linear models perform well with sparse features and lightweight optimization. Future work will focus on validating these findings across larger datasets and exploring adaptive feature–tuning frameworks for real-time fault prediction.



\bibliographystyle{IEEEtran}
\bibliography{ieee-bib}

\end{document}